\documentclass[letter,10pt]{sig-alternate}
\usepackage{cite}
\usepackage{color}
\usepackage{ifpdf}
\ifpdf
\else
\fi

\def\myhyperrefcolor{black}
\usepackage[pdftex,
colorlinks=true,
citecolor=\myhyperrefcolor,
filecolor=\myhyperrefcolor,
linkcolor=\myhyperrefcolor,
urlcolor=\myhyperrefcolor,
pagecolor=\myhyperrefcolor,
linkbordercolor={1 1 1},
bookmarksopen=true,
bookmarksnumbered=true,
frenchlinks=false,
bookmarks=true
]{hyperref}

\usepackage{epsfig,endnotes}
\usepackage{subfigure}
\usepackage{amssymb}
\usepackage{url}
\usepackage{flushend}
\usepackage{tikz}

\newcommand{\DNStamp}{DNStamp}
\newcommand{\domain}[1]{{\tt #1}}
\newcommand{\myparagraph}[1]{{\bf #1~~}}
\newcommand{\etal}{et~al.\ }

\newcommand{\etc}{etc.\ }

\begin{document}

\date{}

\title{\Large \bf DNStamp: Short-lived Trusted Timestamping*}

\author{Christoph Neumannn, Olivier Heen and St\'ephane Onno\\
Technicolor\\
Email: \{christoph.neumann, olivier.heen, stephane.onno\}@technicolor.com
} 

\maketitle

\thispagestyle{empty}

\begin{tikzpicture}[remember picture,overlay]
  \node[rotate=0,scale=1,text opacity=1] at (8, -20)
  {*The extended version of this work is to appear in Elsevier Computer Networks 64 (2014) pp. 208-224.};
\end{tikzpicture}

\begin{abstract}
Trusted timestamping consists in proving that certain data existed at a particular point in time.
Existing timestamping methods require either a centralized 
and dedicated trusted service or the collaboration of other participants using the timestamping service.

We propose a novel trusted timestamping scheme, called \DNStamp, that does not require a dedicated service nor collaboration between participants.
\DNStamp\ produces short-lived timestamps with a validity period of several days.
The generation and verification
involves a large number of Domain Name System cache resolvers, thus removing any single point of failure and any single point of trust.
Any host with Internet access may request or verify a timestamp, with no need to register to any timestamping service.
We provide a full description and analysis of \DNStamp.
We analyze the security against various adversaries and show resistance to forward-dating, back-dating and erasure attacks.
Experiments with our implementation of \DNStamp\ show that one can set and then reliably verify timestamps
even under continuous attack conditions.

\end{abstract}

\section{Introduction}
Trusted timestamping, i.e., proof that certain data existed at a certain time, is indispensable in many situations of the digital world.
Examples of such situations include:
online auctions to ensure correct order of bids;
electronic voting to ensure that the vote was cast at an allowable time;
publication systems to prove that a document was published at a given time and
on-line betting to make sure that a bet is placed before the event.

Trusted timestamping generally relies on trusted and centralized timestamping authorities that provide the timestamp. This introduces a single point of trust, a single point of failure and may limit the scalability of the system. While these constraints are acceptable for some use-cases, they can hinder the functioning or deployment for use-cases that e.g. cannot afford deploying and operating dedicated timestamping servers and infrastructure. 

{\em Temporal decoupling} \cite{froihofer:experience} is a motivating example that requires a trusted timestamping scheme having no single point of failure and being able to absorb peak load. 
Temporal decoupling consists in generating a timestamp proving that a bid or information existed before a given due date. Yet, the bids can be submitted during the validity period of the timestamp, even if the due date has passed.
Overwhelming the server by submitting the bids just before the due date can thus be circumvented. Typical use-cases for temporal decoupling are the submission of tax return or of papers to a conference. Such systems often encounter peak load just before the due date, which could lead to service interruption. 

We propose a completely distributed timestamping scheme, called \DNStamp, that takes advantage of the Internet Domain Name System (DNS). \DNStamp\ involves a large number of DNS resolvers and domain names, thus removing any single point of failure and distributing trust to the DNS resources involved. 
\DNStamp\ provides timestamping for free, i.e., with zero deployment and operational costs, as there is no need to deploy and operate timestamping servers and infrastructure.
In its simplest form, \DNStamp\ is a command line tool. An example timestamp request, followed by the verification is shown below.
\vspace{-0.25cm}

\smallskip
{\footnotesize
\begin{verbatim}
  $ ./dnstamp -request -duration=1day data.7z
  Timestamp set on 2012/01/31-13:22:48
  $ ./dnstamp -verify -date="2012/01/31-13:22:48"
  -duration=1day data.7z
  Timestamp set 02:53:11 hours ago, still valid
\end{verbatim}
}
\smallskip

\begin{figure}
\includegraphics[width=\columnwidth]{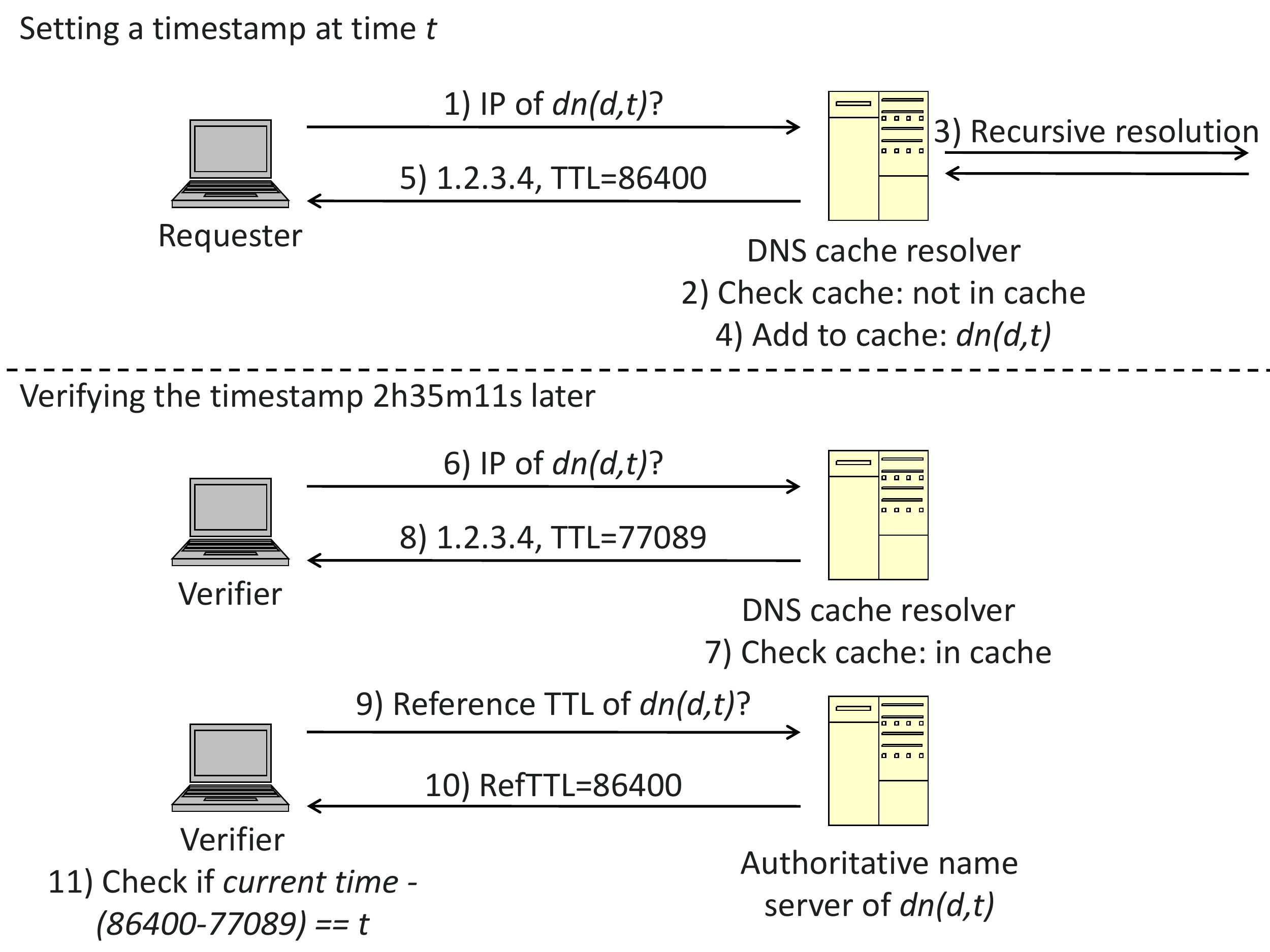}
\caption{Sketch of \DNStamp\ scheme using a single domain name and a single resolver. $dn(d,t)$ denotes a domain name derived from data $d$ to be timestamped and the timestamping time $t$, using a one-way function.}
\label{fig:DNStampoverview}
\end{figure}

\vspace{-0.25cm}
In order to timestamp data with \DNStamp, a requester sends recursive resolution requests to a list of resolvers. The resolution requests contain domain names, which have been derived from the timestamping time (i.e. the current time) and data to be timestamped using a one-way function. Similarly, the list of resolvers is selected using another one-way function. At reception of above resolution requests, the resolvers add new DNS cache entries of the requested domain names into their cache.
The resolvers keep these entries in their cache during a certain Time-To-Live (TTL), a value that is maintained and continuously decremented for each cache entry.
In order to verify the timestamp, typically once data is released, a verifier reiterates the procedure of the requester and retrieves the remaining TTL for each DNS entry.
In addition, the verifier asks for the reference TTL at each domain name's authoritative server. The verifier can calculate the time of the timestamp using the current time, the reference TTL and the remaining TTL. Figure~\ref{fig:DNStampoverview} sketches the \DNStamp\ scheme using a single resolver and a single domain name.

The contributions of this paper are as follows:

(i) We present a novel timestamping scheme called \DNStamp\ that takes advantage of the DNS. 
\DNStamp\ produces short-lived timestamps with a validity period of several days.
Anyone can request and verify timestamps without any dedicated trusted service.

(ii) We formalize the security requirements for trusted timestamping and the associated adversarial model. 
We analyze the security of \DNStamp\ and show its resistance to forward-dating, back-dating and erasure attacks.

(iii) We implement a command line tool of \DNStamp. 
Our experiments from various locations (cloud, broadband access) show that we can reliably set and then verify timestamps. Adversaries with reasonable capabilities cannot overwrite an existing timestamp.

The paper is organized as follows. Section~\ref{ref:DNS} recalls some important concepts of the DNS.  Section~\ref{sec:problemmodel} introduces our security objectives and describes the \DNStamp\ protocol. Section~\ref{sec:DNSdiscussion} discusses the impacts of DNS uses, misuses and security issues on \DNStamp. Section~\ref{sec:analysis} provides a security analysis of \DNStamp.  Section~\ref{sec:implem} presents our implementation and experimental results. Section~\ref{sec:related} presents related work and Section~\ref{sec:conclusion} concludes.

 \vspace{1 cm}
\section{The Domain Name System}
\label{ref:DNS}

\begin{figure}
\includegraphics[width=\columnwidth]{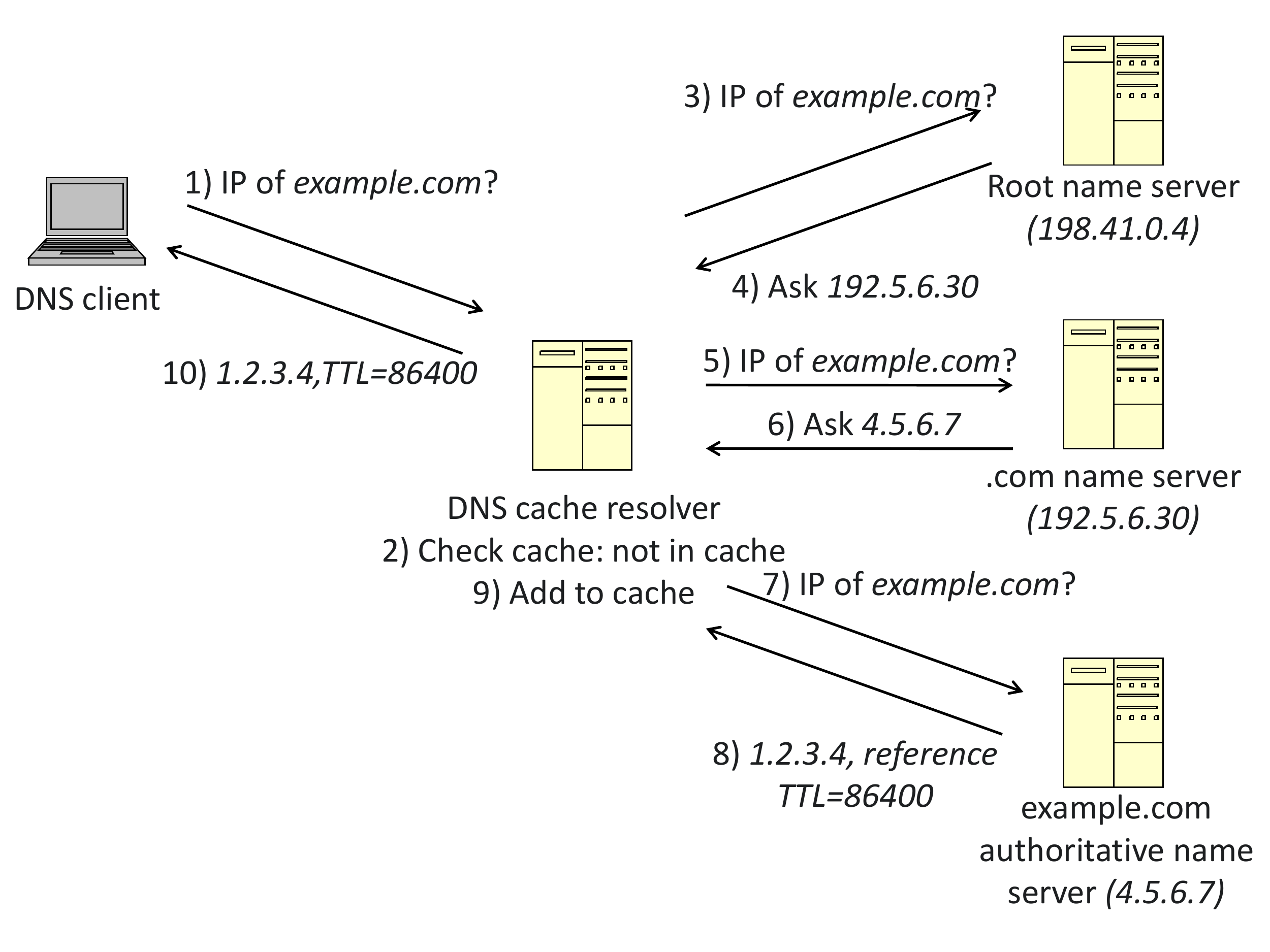}
\caption{Example sequence diagram for a standard DNS resolution request.}
\label{fig:DNSoverview}
\end{figure}
The DNS is a worldwide name resolution service that turns Fully Qualified Domain Names (FQDN), such as \domain{www.example.com}, into the corresponding IP address. This process is called the resolution. 
The DNS also performs the inverse operation, called reverse resolution, which turns an IP address into its FQDN. 

Crucial requirements for the DNS are high-availability and consistency. Therefore, the DNS is highly distributed worldwide, with many local and regional replicas. More precisely, the DNS system relies on a hierarchical structure of DNS name servers. There are currently 13 root name servers responsible for the root zone \domain{`.'}. These root name servers are operated by institutions such as the NASA, ICANN\footnote{\url{https://www.icann.org/}}, VeriSign or RIPE\footnote{\url{https://www.ripe.net/}}. 
Descending one level the DNS hierarchy, a set of regional zones have been defined (\domain{.com}, \domain{.org}, \domain{.net}\ldots), each zone having its own set of name servers. Finally, a particular domain (e.g., \domain{example.com}) has its own name server. A DNS server is said authoritative for a domain if it has the ability to define the corresponding IP addresses and other attributes such as the domain's Time-To-Live. 

To reduce the load on the DNS infrastructure and to increase the overall speed, hosts requesting a DNS resolution do not talk directly to the above authoritative name servers but use a \emph{DNS cache resolver} instead. The resolver is provided typically by the host's ISP.
A resolver maintains a cache of DNS entries. When a resolver receives a resolution request, the resolver first checks whether the requested domain name is in its cache. If the requested domain name is in its cache, the resolver will directly respond. If the entry is not in its cache, the resolver will forward the request to the upper most authoritative DNS name server not in its cache. Once the resolution succeeds, the resolver adds the domain name of the new request (and its intermediate resolution requests) to its cache. Figure~\ref{fig:DNSoverview} depicts an example DNS resolution request.
The host sending a DNS request may specify whether the resolver should forward the request or not. This option is called \emph{recursive resolution}: a recursive resolution request indicates that the resolver should forward the request if the requested domain name is not in its cache; a non-recursive resolution request indicates that the resolver should not forward the resolution request and only respond based on its own cache; the resolver returns an empty response if the requested domain name is not in its cache. 

The replication period of a cache entry in a resolver is called \emph{Time-To-Live (TTL)}. Only an authoritative server can define the \emph{reference TTL} of its domain name. A typical value is 86400 seconds (1 day) but values up to 7 days are supported. A resolver copies the reference TTL when a domain name is added to its cache and afterwards decrements the TTL every second. We call this TTL the remaining TTL.
When answering a resolution request, the resolver returns the remaining TTL. When the remaining TTL expires, the resolver deletes the corresponding cache entry. In order to honor subsequent requests, the resolver must initiate a recursive resolution until the authoritative server is reached.


\section{Timestamping using DNS}
\label{sec:problemmodel}

\label{sec:problem}
Trusted timestamping consists in proving that certain data $d$ existed at a point in time $t_R$. More precisely, a generic timestamping process involves three steps: (i) at time $t_R$, a \emph{requester} timestamps a digest $y=h(d)$ ($h$ is a hash function), which results in the generation of a timestamp $T$, (ii) at time $t_P$, $t_P \geq t_R$, the requester publishes the timestamp $T$ (iii) at time $t_V$, $t_V \geq t_P$, a \emph{verifier}  verifies whether $T$ was actually produced at time $t_R$ as a function of digest $y$.

\begin{table}
\footnotesize{
\begin{center}
\begin{tabular}{|c|l|}
\hline
$d$ & data to be timestamped \\ \hline
$h, h_a, h_b, h_c$ & cryptographic hash functions \\ \hline
$y=h(d)$ & data digest \\ \hline
$t_R$,$t_P$,$t_V$ & request time, publication time \\
&and verification time of a timestamp \\ \hline
$t_A$ & time of an attack \\ \hline
$T$ & a timestamp \\ \hline
$request$ & requesting function \\ \hline
$verif$ & verification function \\ \hline
$\alpha$ & duration of a timestamp \\ 
& in \DNStamp: $\alpha<7 days$  \\ \hline
$[t_R,t_R+\alpha]$ & validity period of a timestamp\\ \hline
$t_R+\alpha$ & expiration time of a timestamp\\ \hline
\end{tabular}
\end{center}
\caption{Notations}
\label{tab:notation}
}
\end{table}

\label{sec:req}


\label{sec:secobj}
We consider a timestamp to be {\em trusted} if it resists to \emph{back-dating}, \emph{forward-dating} and \emph{availability} attacks as follows (Section~\ref{sec:analysis} describes the considered adversaries): 
(i) It should be impossible for an adversary to back-date an existing timestamp $T$. Back-dating consists in changing the date of an existing timestamp $T$ to a time $t_R - \Delta$ prior to $t_R$.
(ii) It should be impossible for an adversary to forward-date an existing timestamp $T$.
Forward-dating consists in changing the date of an existing timestamp $T$ to a time $t_R + \Delta$ after $t_R$. A specific case of forward-dating consists in timestamping using the time $t_A$ at which the attack occurs.
(iii) It should be impossible for an adversary to prevent the request of a new timestamp.
It should be impossible for an adversary to prevent the verification of a valid timestamp.











\label{sec:approach}

\subsection{Requesting a new timestamp}
With DNStamp, a requester generates a list of domain names $D=\langle  dn_1, \ldots, dn_n\rangle$ and a list of resolvers $R=\langle  res_1, \ldots, res_n\rangle$, as described later, which both depend on data digest $y = h(d)$, the generation time $t_R$ and the duration $\alpha$ of the timestamp. Then, she performs the resolution $dn2ip(dn_j,res_j)$ of each domain name $dn_j \in D$ using the resolver $res_j \in R$; she forces recursive resolution in her DNS resolution requests.
As an effect, resolvers that did not cache a requested domain name add this domain name to their cache and set the reference TTL $ref_i$. As specified by the DNS standard, the resolver starts decrementing the TTL and deletes the domain name from its cache when the remaining TTL reaches 0. 
If an entry $dn_j$ was already cached by the resolver $res_j$, the domain reference TTL will not be set, and the resolver will continue to decrement the remaining TTL. The requester can ensure with some probability that a subset of the requested domain names are not already cached, if the sizes of lists $D$ and $R$ are large enough.

The requester generates the above lists $D$ and $R$ as follows.
We suppose that a list of valid resolvers, denoted $rlist$, is provided. 
To compute $D$, the requester generates a list of $m$ IP addresses $ip_1 \ldots ip_m$ using a hash function $h_a$ such that $ip_i = h_a(y||t_R||\alpha||i)$. $||$ denotes the concatenation. She also generates a list of reverse resolvers using a hash function $h_b$ as follows: $inv_i=rlist[h_b(i)]$. She then perform the reverse resolutions $ip2dn(ip_i,inv_i)$ of each $ip_i$ using $inv_i$. The reverse resolution can fail for some IP addresses, since not all possible IP addresses have a corresponding domain name. The result is the list of domains $D=\langle dn_1, \ldots, dn_n\rangle$ with $n < m$. We denote $I$ the list of indices $j$ such that the reverse resolution of $ip_j$ returns a valid domain name. 
$R$ is computed such that $res_i=rlist[h_c(y||t_R||\alpha||i)]$ where $i \in I$.

The resulting timestamp is $request(y||t_R||\alpha)=T=(D,I)$.

\subsection{Publishing the timestamp}
At time $t_P$, the requester publishes $T$ to the world (web) or to a group (social network), or directly to some verifiers (email). We also recommend publishing $y$ and $t_R||\alpha$ along with the timestamp $T$ (see Section~\ref{sec:analysis}).

\subsection{Verifying an existing timestamp}
\label{sec:verif}
We suppose that the verifier retrieved $T$, $y$, $t_R$ and $\alpha$. 
The verifier computes the list of resolvers $R$ using the same algorithm than the requester.
She then retrieves the remaining TTL for each domain name $dn_j \in D$ using the resolvers $res_j \in R$.
In addition, she retrieves the domain reference TTL for each domain name in $D$ by querying the respective authoritative servers (see details in Section~\ref{sec:implem}).
The verification succeeds if, for a portion $1-\epsilon$ of the domain names of $D$, the difference between the domain reference TTL ($ref_i$) and the remaining TTL $(rem_i)$ is consistent with the time provided with the proof, i.e. $t_V = t_R + (ref_i - rem_i)$.
Finally, having $y$, $t_R$ and $\alpha$, she can also generate the list of domain names $D$ and verify if it is equal to the one contained in $T$. The above procedure corresponds to checking whether $verif(request(y||t_R||\alpha))=T)=t_R$.




\section{Subtleties of the DNS}
\label{sec:DNSdiscussion}
The primary purpose of the DNS is to translate domain names into IP addresses. However, over the years many different uses and security issues appeared.
This section presents non-malicious uses and implementation specificities as well as malicious uses and security issues of the DNS and discusses their impact on \DNStamp. Finally, this section presents two variants of \DNStamp.

\subsection{Non-malicious uses and specificities}
The DNS is sometimes used as a load-balancing mechanism. One technique, called round-robin DNS, returns a list of IP addresses instead of a single IP address. The host that sent the resolution request may pick any of the returned IP addresses. Another technique, used by Content Delivery Networks such as Akamai, sets short TTL values ranging from a few seconds to several hours \cite{Dilley2002}. This enables dynamic redirection of the domain names to other servers and locations as condition changes.
Content delivery networks also localize the provenance of a DNS resolution requests (generally based on the resolver's IP address) in order to return the IP addresses of servers close to the requester \cite{Ager2010}.
These optimizations are not an issue for \DNStamp. The domain selection algorithm of \DNStamp\ (see details in Section~\ref{sec:domainselection}) removes all TTL below the expected duration of the timestamp. This eliminates small TTL values used by content delivery networks. The usage of round-robin DNS is also not an issue, since round-robin adds multiple entries into resolver, all set to the same TTL. A subsequent DNS resolution request on the same domain returns the same list of entries, which can all be used to calculate the timestamp. Finally, the localization optimizations used by content delivery networks are not an issue since the resolver returns its cache entries independently of the IP address of the verifier.

Some resolvers use load-balancing themselves to support a large load of resolution requests. If properly configured, the servers involved in load-balancing share the same cache. However, some resolvers use load-balancing with poor cache sharing \cite{Ager2010}.
As a consequence, recursive resolution requests generate additional traffic even if a corresponding DNS entry is already in the cache of one the servers.
This abnormal behavior can impact \DNStamp\ as the resolvers may return fresh cache entries (thus with bad TTL) when verifying a timestamp. Two techniques allow to mitigate this issue: (i) distributing a single timestamp over a large number of resolvers reduces the impact of badly configured resolvers and (ii) requesting a domain name several times in a row during the timestamp request ensures that the entry is added to most servers involved in load-balancing.

Several companies\footnote{e.g. Spamhaus \url{http://www.spamhaus.org}} provide DNS-based blackhole lists (DNSBL) to test whether a domain name or IP-address is known to be malicious or not \cite{Ramachandran2006a,Ramachandran2006b}. To verify a domain \domain{test.net} a client can send a DNS resolution request such as \domain{test.net.dnsbl.example.com}, where \domain{example.com} is the DNSBL service provider. The returned IP address, typically a loopback address, indicates whether the domain is blacklisted or not. 
DNS-based blackhole lists have no impact on \DNStamp\ that uses random publicly addressable IP addresses to select a domain name. This ensures that domain names of the type \domain{test.net.dnsbl.example.com} are not selected.
\smallskip


We noticed unexpected variations in the handling of \domain{CNAME} records by resolvers.
A \domain{CNAME} record is an alias towards another domain name.
For instance, the alias \domain{www.linux.org} points to the classical \domain{A} record \domain{linux.org}.
Following the DNS specifications, resolvers must keep independent TTL, one for the \domain{CNAME} record and one for the \domain{A} record.
However, it appears that a proportion of resolvers do not follow this expected behavior.
Instead, they do reset the TTL of the \domain{A} record when they resolve the corresponding alias.
This behavior is revealed in the session hereafter.
\vspace{-0.2cm}

\smallskip
{\footnotesize
\begin{verbatim}
  $ dig @24.180.22.42 +noall +answer linux.org
  linux.org.      14400  IN  A      209.92.24.80
  $ dig @24.180.22.42 +noall +answer linux.org
  linux.org.      14392  IN  A      209.92.24.80
  $ dig @24.180.22.42 +noall +answer www.linux.org
  www.linux.org.  14400  IN  CNAME  linux.org.
  linux.org.      14400  IN  A      209.92.24.80
  $ dig @24.180.22.42 +noall +answer linux.org
  linux.org.      14398  IN  A      209.92.24.80
\end{verbatim}
}
\smallskip

\vspace{-0.2cm}
A weak attacker could use these "resetters" against \DNStamp\ as follows: the attacker guesses as much as possible \domain{CNAME} records that point towards records used by a timestamp. The attacker then resolves the \domain{CNAME} records and, if resetters are frequent enough, modifies a significant portion of the timestamp. To evaluate the impact of such an attack, we evaluate the number of resetters.
From the resolver list $rlist$ we select a sample of 4469 resolvers that answer plausible version information\footnote{we use the command {\tt dig @IP version.bind txt chaos}.}.
We then test how each resolver handles \domain{CNAME} records and get the following results:
87.8\% do not reset, which is the expected behavior; 5.6\% announce a reset but do not actually reset, which is harmless for \DNStamp; 4.3\% do actually reset, which is the bad case for \DNStamp; 2.3\% fall in less significant cases.
In light of the above discussion, a realistic countermeasure consists in removing all resetters (4.3\%) from $rlist$.


\subsection{Malicious uses and security issues}
The DNS is also used for malicious purposes. Similarly to content delivery network, fast-flux service networks rely on DNS round-robin and short TTL \cite{Holz2008a,Castelluccia2009}. Fast-flux is used by malware to hide the location of malicious servers. It consists in constantly changing (through short TTL) the list of IP-addresses being associated to the malicious domain name. The IP addresses are selected from a large pool of compromised machines, which act as proxies and redirect the connection to the malicious server.
\DNStamp\ can cope with fast-flux exactly the same way \DNStamp\ copes with content delivery networks.

Some domain names are specifically created for malicious activity, e.g. for spam campaigns, for a fast flux service network or for a botnet. These domain names have very specific characteristics which can include short TTL, high number of distinct IP addresses, short domain life time (e.g. suddenly disappearing after several day of existence), the string of the domain name (e.g. in the English dictionary or not), abrupt lookup patterns \etc \cite{Hao2011,Bilge}. 
Attacks exist to maintain such domains in caches even if the domain has been revoked by name servers which are higher in the DNS hierarchy \cite{Jiang}.
\DNStamp\ may use domain names used for malicious activities. These domain names are maintained in caches as other cache entries and their TTL decrements normally. \DNStamp\ also checks a set of criteria during the domain name selection, typically sufficient high TTL values (see Section~\ref{sec:domainselection}), which ensure that the domains can be used for timestamping.

\domain{NXDOMAIN} hijacking is another malicious use of the DNS: some resolvers return specific IP addresses when an empty domain name has been requested, instead of the \domain{NXDOMAIN} message \cite{Ager2010}. Their objective is to gain some money by pointing to IP addresses of search engines or advertisement-websites. In \DNStamp, the domain name selection process is based on IP-addresses and performs checks to eliminate \domain{NXDOMAIN}s (see Section~\ref{sec:domainselection}).

Cache poisoning \cite{Kaminsky2008} attempts to inject false cache entries into resolvers. The objective is to associate a legitimate domain name to an IP address that the attacker controls. The attack was possible because some fields in a DNS request were predictable, and a malicious name server could respond in lieu of the authorized name server. Several mitigations have been proposed since \cite{Steinho2006, Herzberg2012}. 
\DNStamp\ is affected by the cache poisoning attack. An attacker may try to inject false TTL values into the cache using cache poisoning.
By distributing a single timestamp over a large number of domain names and resolvers, we reduce the impact of an attacker that could poison a vulnerable resolver.
Still, a long term solution to cache poisoning (for the DNS and for \DNStamp\ as well) would consist in moving to DNSSEC.

Finally, DNS reflector attacks \cite{Paxson:2001:AUR:505659.505664} use open DNS cache resolvers as traffic amplifiers during a DoS attack. The existence of open DNS cache made this attack possible. \DNStamp\ does neither increase nor decrease the importance of this attack.


\subsection{Variants of \DNStamp}
\DNStamp, as described in Section~\ref{sec:approach} does not work with IPv6.
The domains $\langle dn_1, \ldots, dn_n\rangle$ are obtained from the inverse resolution of IPv4 addresses.
This method succeeds because i) almost all IPv4 addresses are assigned, and ii) enough IPv4 addresses reverse to a valid domain name (our experiments report {\raise.17ex\hbox{$\scriptstyle\sim$}}$20\%$).

With IPv6, the first condition does not hold:
the probability of a random IPv6 address being assigned is close to 0 due to the size of the IPv6 address space. 
A solution then is to choose $\langle dn_1, \ldots, dn_n\rangle$ among short second level domain names, typically three to four characters long.
This method succeeds if (i) the probability of finding valid domain names is high, and (ii) the number of domain names is large enough.
We exhaustively tested four ways of constructing random domain names as depicted in Table~\ref{tab:ipv6}.
Our measurements indicate that short domain names generate a sufficient number of valid domain names to be used for \DNStamp.

\begin{table}
\begin{center}
\begin{tabular}{|l|c|c|}
\hline
Domain	& \# & probability	\\\hline
{[A-Z0-9]\{3\}.com}	& 46656 & 98.1\%\\
{[A-Z]\{4\}.org}	& 456976 & 32.9\%\\
{[A-Z]\{4\}.com}	& 456976 & 97.5\%\\
{[A-Z0-9]\{4\}.com}	& 1679616 & 38.1\%\\\hline
\end{tabular}
\end{center}
\caption{Observed probabilities of finding valid short domains.}
\label{tab:ipv6}
\end{table}




We now introduce another variant of \DNStamp\ with interesting characteristics such as generating timestamps with a long validity period. Yet, this variant introduces a single point of trust.
We suppose that we own a given domain \domain{example.com} and control the corresponding authoritative server. Instead of randomly choosing IP addresses, we generate a list of strings $\langle s_1 \ldots s_n\rangle$ that depend on data digest $y=h(d)$ and $s$. The resolution request occurs against the domains $s_1$\domain{.example.com}, \ldots $s_n$\domain{.example.com} using the list of resolvers $R$ as described and computed before. The authoritative server resolves each string generated using a predefined TTL. 
Dedicating an authoritative server to timestamping enables to set reference TTL values higher than the generally observed 7 days or less. The standard \cite{rfc2181} specifies the TTL as a 32 bit unsigned number ranging from 0 to 2147483647 seconds, thus about 68 years. We observed in our experiments that about one third of the resolvers set a remaining TTL of 1 day even if the reference TTL is higher (see Section~\ref{sec:moredays}). We can expect that such behavior becomes more important with significantly higher TTL values. A solution would consist in using resolvers dedicated to the timestamping service, or in using only certified resolvers supporting high TTL values.

\section{Security Analysis}
\label{sec:analysis}
We first define the considered adversaries. Then, we discuss attacks against the timestamping process and we consider forward-dating and back-dating attacks. We discuss the specific case of a very strong adversary and finish with attacks against the timestamping software itself and privacy considerations.

\subsection{Adversary models}
\label{sec:adversaries}

A \emph{weak adversary} has the same capabilities as a regular verifier. We suppose that once a timestamp has been set the weak adversary has knowledge of $T$, $y$, $t_R$ and $\alpha$. A weak adversary can send any number of regular DNS requests to any DNS server.
She cannot eavesdrop the traffic generated by a timestamp request nor by a timestamp verification.
She has no particular privilege or right over the DNS. She has no control over any particular resolver or authoritative server.

An \emph{intermediate adversary} has the same capabilities as a weak adversary. Additionally, the intermediate adversary can eavesdrop all messages between the requester and any DNS server, and between a verifier and any DNS server.

A \emph{strong adversary} has the same capabilities as an intermediate adversary. Additionally, the strong adversary controls a limited number of resolvers and authoritative servers. 

A \emph{very strong adversary} can eavesdrop, alter, replay or inject any message between the requester and the DNS. This adversary is equivalent to a Dolev-Yao adversary \cite{Dolev:1981:SPK:891726} that controls the entire network.

\subsection{Attacking the timestamping process}
\label{sec:algoVerif}

\myparagraph{Early disclosure of a timestamp}
In this attack, an intermediate adversary tries to deduce a timestamp before it is disclosed by the requester.
The intermediate adversary observes all calls to the $ip2dn$ function during the request of a timestamp.
The intermediate adversary records all reverse resolution requests that are grouped in a very short period of time and that have the same source address (i.e. the IP address of the requester).
This provides a superset of values $\langle dn_1,...,dn_n\rangle$ that will be used by the requester to set the timestamp.
The intermediate adversary performs non-recursive resolutions for each captured $dn_i$ to all resolvers in the list $rlist$.
Whenever the domain $dn_i$ appears to be in the cache of the DNS resolver $res_j$, the intermediate adversary deduces that $dn_i$ was used by the requester to set the timestamp.
Doing so, the intermediate adversary ends up learning all the elements of the timestamp: the observed domain names $dn_i$, the deduced index numbers $I_i$, the deduced resolvers $res_j$.  The intermediate adversary can also deduce the probable duration of the timestamp: this duration is majored by the minimum TTL observed in domains $dn_i$.
The intermediate adversary is then able to disclose this information {\em before} the requester.
Note that, the adversary is {\em not} able to disclose the digest $y$ or data $d$ a fortiori.
However, a careless verifier may successfully verify the early disclosed timestamp.
Later on, once $y$ or $d$ is published, the verifier may believe that the adversary really made the timestamp.

This attack typically requires $|rlist|.m$ resolutions from the adversary, where $m$ is the number of inverse resolutions performed by the requester.
With the current implementation of \DNStamp, we have $|rlist|=11000$ and $m=1100$ for a typical 1 day timestamp.
Thus, the intermediate adversary must perform $12,100,000$ resolutions.
This value is high but still tractable for an intermediate adversary having a lot of bandwidth.
We recommend that all the elements of the timestamp $T$ be considered public as soon as the requesting process starts.
In particular, we recommend avoiding use cases where the requester starts a request process and then waits for long before publishing the timestamp $T$.

\myparagraph{Early setting of a timestamp}
In this attack, an intermediate adversary tries to set a timestamp {\em before} the requester has completed the timestamping process.
The adversary observes the $dn_i$ during the requester's calls to the $ip2dn$.
She can then set the TTL for $dn_i$ in all $res_j$ before the requester sets the TTL for $dn_i$ in some $res_k$.
As a result, the cache of $res_k$ will expire a bit earlier than if set through the un-attacked process.

We argue that this is not a serious threat.
First, the intermediate adversary must write all cache entries before the requester does.
Even though, the anticipated timestamp will precede the expected timestamp by a short duration. This duration corresponds to the delay before the requester writes the cache entries.
Because we use the value $t_R$ for the generation of the IP addresses $ip_i$, a careful verifier can detect the attack.
The time of the timestamp will be earlier than the time $t_R$, instead of being equal.

\myparagraph{Disturbing a timestamp request}
A very strong adversary can easily block a timestamp, this case is discussed in Section \ref{ssec:verystrong}.
An intermediate adversary cannot rewrite DNS traffic, but she may try to disturb the timestamping process.
By performing the observations of the ''early disclosure attack'', the intermediate adversary can overload the resolvers as soon as they are selected by the requester to set the timestamp.
This gives partial control over the function $dn2ip$.
If these resolvers are overwhelmed, they will not fulfill the recursive request.
As a result some $(dn_i,i)$ pairs in the timestamp $T$ will not have the domain $dn_i$ set in a resolver.
The verifier of $T$ will get less corresponding pairs, possibly reaching a threshold above which the timestamp is destroyed.
As a countermeasure, the requester may retry the recursive requests if it was unsuccessful.
This raises the cost of overloading the resolvers for the intermediate adversary.

\myparagraph{Disturbing a timestamp verification}
An intermediate adversary may disturb the resolvers that will be used during a verification process.
This attack proceeds in a fashion similar to the former attack, but is only executed if a timestamp has already been set.

\subsection{Forward-dating and back-dating attacks}

This section analyses if an adversary can forward-date or back-date a timestamp. We suppose that a timestamp $T$ has already been requested and set.
During verification, the timestamping time is calculated according to $t = t_V - (ref - rem)$. We consider that it is up to the verifier to ensure that its local time $t_V$ is correct. An adversary may thus try to modify $ref$ or $rem$ to forward-date or back-date a timestamp. 

If an adversary were able to override the remaining TTL $rem$ of an existing cache entry with a TTL $rem_A = rem + \Delta > rem$, then she would be capable of forward-dating a timestamp.
Similarly, if an adversary were able to override the remaining TTL $rem$ of an existing cache entry with a TTL $rem_A = rem - \Delta < rem$, then she would be capable of back-dating a timestamp.
The attacked timestamp is $t_V - (ref - (rem \pm \Delta)) = t_R \mp \Delta$.

If an adversary were able to override the reference TTL $ref$ of a domain name entry with a TTL $ref_A = ref - \Delta < ref$, then she would be capable of back-dating a timestamp.
Similarly, if an adversary were able to override the remaining TTL $ref$ of a domain name entry with a TTL $ref_A = ref + \Delta > ref$, then she would be capable of back-dating a timestamp.
The attacked timestamp is $t_V - ((ref \pm \Delta) - rem) = t_R \mp \Delta$.

\label{sec:forward}
\label{sec:backdate}

\myparagraph{Weak and intermediate adversary}
The weak and intermediate adversaries have the same capabilities regarding the $rem$ and $ref$ values. The weak and intermediate adversary cannot change $ref$, since they cannot modify the answers of an authority server or otherwise tamper the status of an authoritative server. 
The weak and intermediate adversary may try to modify $rem$ by sending resolution requests to resolvers. 
According to the DNS standard \cite{Mockapetris1987}, the resolver does not update the $rem$ value of a cached domain name if a resolution request for this domain occurs. Instead, the resolver simply returns the cached answers with a TTL $rem$ and decrements the value $rem$ by 1 every second as usual, as long as $rem > 0$. We have $rem > 0$ for all domain names until the expiration time $t_R + \alpha$ of the timestamp.
Thus, if all resolvers comply with the standard, an adversary cannot forward-date nor back-date an existing timestamp until its expiration time $t_R + \alpha$. Our experiments confirm that it is not possible for a weak adversary to overwrite the remaining TTL of an existing cache entry during a timestamps lifetime (see Section~\ref{sec:overwrite}). 

After $t_R + \alpha$, back-dating is still not possible, since a weak or intermediate adversary can only reset a cache entry to the reference TTL. As we will demonstrate below, after $t_R + \alpha$, forward-dating is also not possible if $T$ depends on parameters $t_R$ and $\alpha$. We first show that there exist a forward-dating attack if $T$ does not depend on $t_R$ and $\alpha$. Then, we show that the attack is not possible anymore if $T$ depends on parameters $t_R$ and $\alpha$.

\myparagraph{Vulnerable case $T=request(y)$:}
Once a timestamp $T$ has expired, an adversary possessing $T$ may set a new timestamp that replaces the previous one such that $verif(T)=t_A > t_R + \alpha$. The adversary does not need $y$ or $d$ to perform this attack. The adversary simply requests a recursive resolution of all the domain names of $dn_i \in D$ using the resolvers $res_i \in R$. The resolvers do not cache $dn_i$, thus the caches are updated and the requested domain names $dn_i$ are cached with the reference TTL.
At verification time $t_V$, the verifier does not know whether the genuine timestamp $T$ has already expired or not (i.e. if $t_V$ greater or smaller than $t_R + \alpha$). Thus in general, the verifier does not know whether the returned time $t=verif(T)$ is $t_R$ or $t_A$, and therefore cannot trust the returned time. 

\myparagraph{Non-vulnerable case $T=request(y||t_R||\alpha)$:}
As long as $y$, $t_R$ and $\alpha$ have not been released by the requester, the verifier does not know whether the timestamp $T$ has expired or not; the function $verif(T)$ returns a timestamp and the verifier has no means to verify if this timestamp has been modified by an adversary or not. 
If $t_R$ and $\alpha$ have been disclosed, the verifier knows whether the timestamp has expired or not and can check that $verif(request(y||t_R||\alpha)=T)=t_R$.
Once a timestamp $T$ has expired, an adversary possessing $T$ may replace the domain name cache entries  $dn_i \in D$ using the resolvers $res_i \in R$ and set a new time $t_A$. However, the verifier will detect the attack because $verif(request(y||t_R||\alpha)=T) \neq t_R$.
An adversary may of course request its own timestamp $T'$ using $t_A$; however, this does not attack the timestamp $T$.

\myparagraph{Strong adversary}
A strong adversary is capable of changing the $ref$ values for the authoritative server she controls.
Similarly, a strong adversary is capable of changing the $rem$ values for a resolvers she controls.

If \DNStamp\ exclusively relied on domains and resolvers controlled by a strong adversary, the adversary would be able to forward-date and back-date any timestamp. To address this issue, \DNStamp\ distributes its trust over a large number of domain names and resolvers, typically 100 as in our experiments in Section~\ref{sec:implem}. An adversary needs to control a high number of domain names and resolvers to back-date the timestamp.

We depict the effect of above strategy with an example. For one timestamp we suppose that we require 100 domain names to be stored in 100 different resolvers. We validate a timestamp if more than $50\%$ of these entries return the same timestamp.
According to the hypergeometric distribution, a strong adversary needs to control 4194 resolvers of the 11k resolvers in $rlist$ in order to modify a given timestamp with a probability $1\%$. Similarly, an adversary that controls $15\%$ of the 11k resolvers attacks a given timestamp with a probability of $2.2*10^{-16}$.
Finally, to depict the required power of an adversary that controls some domains, we assume that our algorithm randomly chooses domains out of the {\raise.17ex\hbox{$\scriptstyle\sim$}}143 million active domains.\footnote{Numbers retrieved from DomainTools, \url{http://www.domaintools.com/internet-statistics/}, Retrieved Jan. 2013.} An adversary would need to control $21$ million in order to modify a given timestamp with a probability $1.1*10^{-16}$.






\subsection{The case of a very strong adversary}
\label{ssec:verystrong}
The very strong adversary can back-date, forward-date, block any timestamp. To forward or back-date a timestamp, the very strong adversary modifies the remaining or the reference TTL of the resolver responses. To block a timestamp, the very strong adversary drops all related DNS requests or responses.

A very strong adversary can forward or back-date a timestamp even if DNSSec \cite{RFC4033,RFC4035} is used. DNSSec provides integrity and authenticity for DNS records returned by a resolver. In particular, DNSSec ensures the integrity of the reference TTL. A very strong adversary cannot modify the reference TTL without altering the integrity of the resolver responses. However, DNSSec does not protect the remaining TTL, and the 
very strong
adversary can set any arbitrary remaining TTL without being detected as long as the modified TTL is smaller than the reference TTL.

One classical workaround is the use of a tunnel towards an Internet application server.
For instance, an SSL or SSH server performs \DNStamp\ requests on behalf of the requester.
This solution may be useful, but breaks our requirement of no single point of trust.

Thus, \DNStamp\ is not resistant to a very strong adversary. A typical very strong adversary is an Internet Service Provider that performs deep packet inspection on all its traffic. A very strong adversary can also be a malware running on the host of the requester or verifier and inspecting all outgoing traffic.


\subsection{Attacking the timestamping software}
\label{sec:softwarattack}
As any piece of software, \DNStamp\ may be altered.
If unnoticed, such alteration would 
invalidate existing timestamps or redirect timestamps to specific domain names or resolvers.
For instance, the domain name selection can be changed so that the software preferably selects domain names controlled by the attacker. Similarly, the resolver list $rlist$ can be modified towards controlled resolvers. 
To mitigate above issues, classical software protection mechanisms should be implemented, such as providing a hash of both the \DNStamp\ software and $rlist$.


\subsection{Privacy considerations}
\label{sec:privacy}
Privacy is not one design goal of \DNStamp.
However, we conjecture that once a timestamp is properly set, a weak adversary cannot deduce participant information.
In particular, she cannot link timestamps to a requester, or tell when a timestamp was verified and by whom.
This is due to \DNStamp\ not requesting identity related information and resolvers not revealing the sources of DNS operations.
Intermediate adversaries and above probably break many privacy properties as they directly observe the full DNS traffic from the requester.


\section{Measurements and Experiments}
\label{sec:implem}

We have implemented DNStamp by extending and adapting the EphPub command-line tool \cite{Castelluccia2011}. 
The prototype takes as input the file to be timestamped, the duration the timestamp should stay valid and an option indicating whether to request or verify the timestamp. Our prototype works with IPv4 addresses only and supports a timestamp duration up to 1 day. The case of timestamps
different than 1 day is discussed in Section~\ref{sec:moredays}.

\subsection{Selecting domain names}
\label{sec:domainselection}

In our experiments we require that each timestamp generates $100$ resolution requests. We need to generate $100$ random domain names with a reference TTL equal to 1 day. We do so by executing the domain name selection process described below.

First, we generate a list of random IP addresses and try a reverse resolution for each of them. The reverse resolution fails for an important number of IPv4 addresses. This is normal, since not all of the possible IPv4 addresses have an assigned domain name. In addition, we require that (i) the returned domain name is valid, i.e. that it does not generate an \domain{NXDOMAIN} error \cite{Mockapetris1987} when queried and that (ii) an authoritative server exists for the returned domain. Our measurements over 100 timestamping processes shows that on average $20\%$ of the randomly generated IPv4 can be successfully reversed to a valid domain name.

Then, we retrieve the reference TTL for each domain name and verify whether the reference TTL is strictly equal to 1 day. We drop a domain name if it does not satisfy this constraint. We continue to search for domains that satisfy this constraint until we reach the $100$ expected domain names.
To retrieve the reference TTL for a given domain, we emit an \domain{SOA} query \cite{Mockapetris1987}. This returns the authoritative server of the domain name and related information such as the \domain{minimum TTL}. According to the standard, this TTL should be used as the reference TTL. However, we observed that this was not always the case. Therefore, we decided to use a more accurate source of information for the reference TTL consisting in sending a DNS resolution (\domain{A} query) to the authoritative server itself. 

\begin{figure}
\begin{center}
\includegraphics[width=\columnwidth]{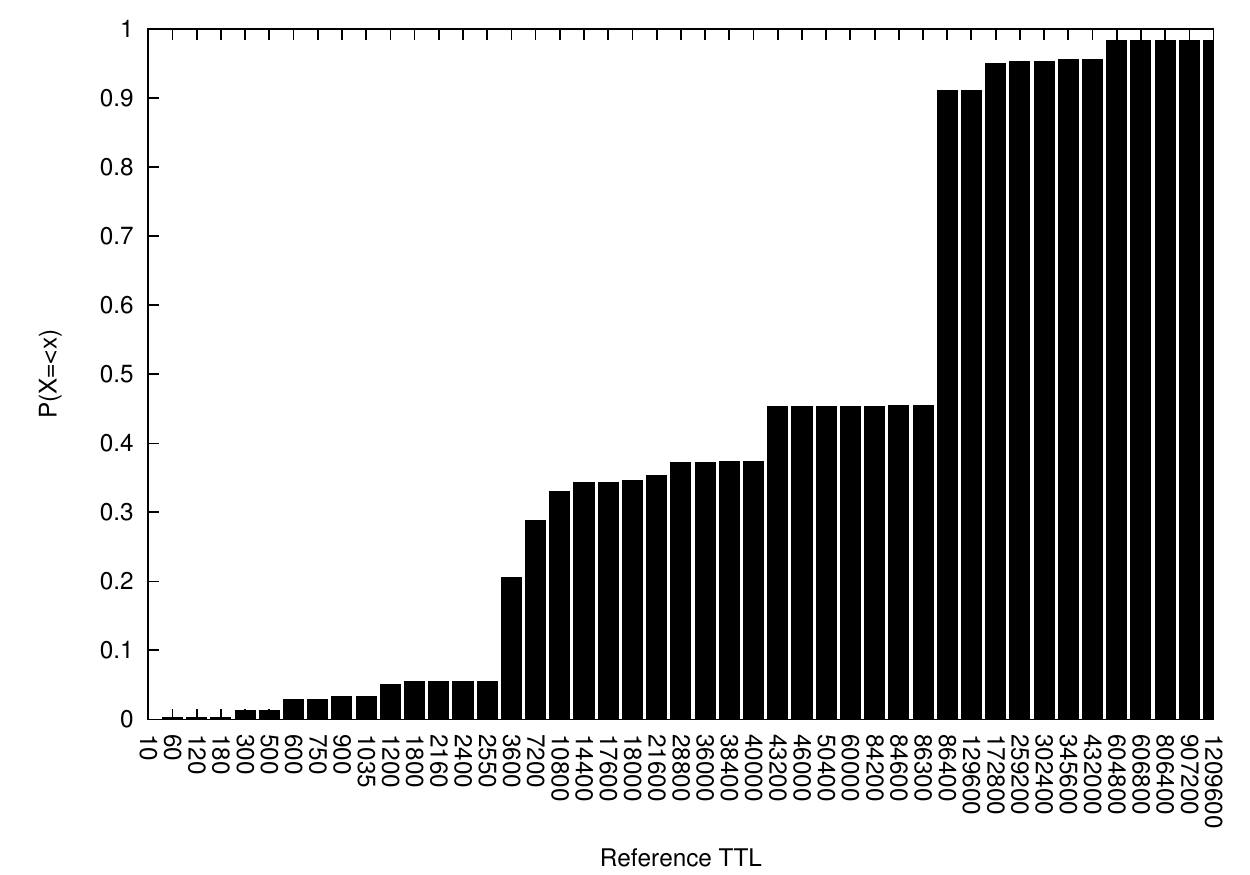}
\caption{CDF of reference TTL of domains observed during timestamp validations, representing a total of {\raise.17ex\hbox{$\scriptstyle\sim$}}757K domain names.}
\label{fig:distribTTL}
\end{center}
\end{figure}

Figure~\ref{fig:distribTTL} shows the distribution of reference TTL we observe during our experiment. This distribution is consistent with the results presented by \cite{Sit2002}. According to this distribution, if the targeted reference TTL is equal to 1 day, we must drop on average $54\%$ of the returned domain names.

\begin{figure*}
\begin{center}
\subfigure[]{
\includegraphics[width=\columnwidth]{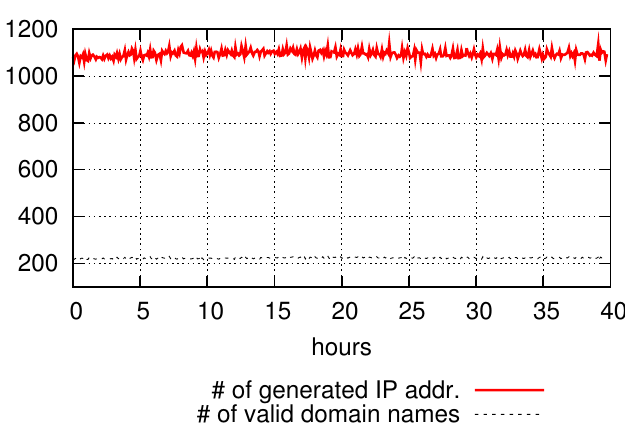}
\label{fig:domainGen}
}
\subfigure[]{
\includegraphics[width=\columnwidth]{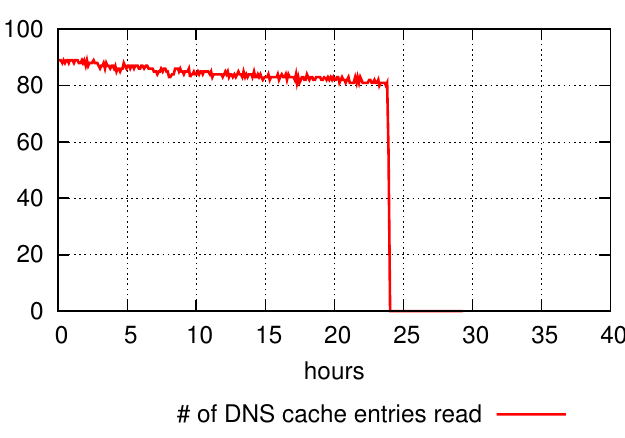}
\label{fig:ReadTS}
}
\caption{\subref{fig:domainGen}  IPv4 addresses and valid domain names retrieved per timestamp verification to generate 100 valid domain names with 1 day TTL. \subref{fig:ReadTS}  Cache entries read.  Numbers are averages from five 1 day timestamps $T$ continuously verified during 40 hours.}
\label{fig:TSgeneration}
\end{center}
\end{figure*}

Figure~\ref{fig:domainGen} shows that on average our prototype tests $1100$ random IP addresses to generate $100$ domain names with a TTL equal to 1 day. Approx.\ $220$ IP addresses are successfully reversed to a domain name that is valid and has an authoritative server.  These numbers stay constant over time.

\subsection{Load and delay considerations}
The important number of IP addresses and domain names that needs to be generated and tested for one timestamp could seem prohibitive. Our implementation requires about $3$ minutes to iterate through and test all IP addresses and domain names and to eventually set the timestamp. We believe that this delay comes from our unoptimized implementation and that we can reduce the timestamping delay: 
(i) by reducing the number of cache entries required for one timestamp, and (ii) by dispatching and multi-threading batches of IP reverse-resolutions and search of authority requests among a larger set of DNS server.
This delay generates a timestamp that is shifted compared to $t_R$. The requester may estimate and add this delay to $t_R$ before selecting the domain names.
Otherwise, verifiers will observe the shift and would have to tolerate a delay representative of a requesting time.

The verification process does not need to reiterate through all IP addresses and domains, thus generating less load than the timestamp request. The verifier may directly use the domain names and resolvers described in the structure $T$.
Once the timestamping time has been retrieved, the verifier may test whether the domains and resolvers are actually bound to the timestamped data $d$ by executing the domain selection algorithm.

\begin{table}
\footnotesize{
\begin{center}
\begin{tabular}{|l|c|c|c|}
\hline
Experiment & Duration & Requests & Approx. Load \\
	& (sec.) & & (requests/sec.) \\ \hline
1 hour \DNStamp & 94 & 1954 & 20.8 \\ \hline
2 hours \DNStamp & 154 & 2214 & 14.4 \\ \hline
1 day \DNStamp & 189 & 3564 & 18.8 \\ \hline
2 days \DNStamp & 1836 & 25911 & 14.1 \\ \hline
Start surf & 12 & 94 & 7.8 \\ \hline
Regular surf & 600 & 2543 & 4.1 \\ \hline
Lab activity & 9 hours & 207377 & 6.4 \\ \hline
\end{tabular}
\end{center}
\caption{Comparison of DNS loads}
\label{tab:load}
}
\end{table}

We also evaluate the additional load of \DNStamp\ on the Domain Name System compared to other DNS consuming activities. Table~\ref{tab:load} summarizes the experimental results. With the current implementation of \DNStamp\ a request generates a peak load of 15 to 20 DNS requests per second. 
The experiment called start surf consists in starting the Internet browser Google Chrome and searching for the term "timestamping". The experiment called regular surf consists in consulting a new webpage per minute during 10 minutes.
The experiment called lab activity consists in recording the DNS activity of a 65 hosts lab during 9 working hours. 
\DNStamp\ generates a higher DNS peak activity than regular surf, yet it remains in the same order of magnitude.

\subsection{Selecting resolvers}
\label{sec:selectDNSres}
The selection of resolvers relies on a list of publicly accessible resolvers.
For our experiments, we use a list of {\raise.17ex\hbox{$\scriptstyle\sim$}}11k resolvers.
We obtained this list by keeping the resolvers that were still online at the time of the experiment from the list of {\raise.17ex\hbox{$\scriptstyle\sim$}}22k resolvers provided by EphPub \cite{Castelluccia2011}.
The latter list dates back to November 2009.
This highlights that the list of resolvers needs to be maintained and continuously updated: old resolvers may go offline and new ones may be added.
The methods for constructing the resolver list may rely on the methods proposed by \cite{Dagon2008} and \cite{Castelluccia2011}. Castelluccia \etal \cite{Castelluccia2011} estimated the number of resolvers that perform caching properly to 1.7 million.
As discussed in Section~\ref{sec:softwarattack}, the list of resolvers has to be considered as an intrinsic part the timestamping software. Updating this list is equivalent to changing the domain selection process. If the list changes, the resolvers selected for a particular timestamp changes, thus making existing timestamps invalid.

\subsection{Retrieving a timestamp}
Figure~\ref{fig:ReadTS} shows the number of cache entries our prototype could successfully read while continuously retrieving a timestamp over 40 hours. The timestamp completely vanishes after 24 hours. This behavior is normal since all remaining TTL reach 0 after 24 hours.

We observe that about $90\%$ of the cache entries can be read right after the timestamp has been set. The missing $10\%$ correspond to the cache entries that could not be successfully set by the requester. Indeed, we noted that some resolvers either refuse a connection (returning a DNS \domain{REFUSED} message) or simply time out. This ratio slightly decreases, and after $20$ hours and until the expiration of the timestamp, about $80\%$ of the cache entries are successfully retrieved. The decrease of valid cache entries typically comes from cache leaks \cite{Kumar1993}. Still, the remaining fraction of valid cache entries is sufficient to validate the timestamping time.

\begin{figure}
\begin{center}
\includegraphics[width=\columnwidth]{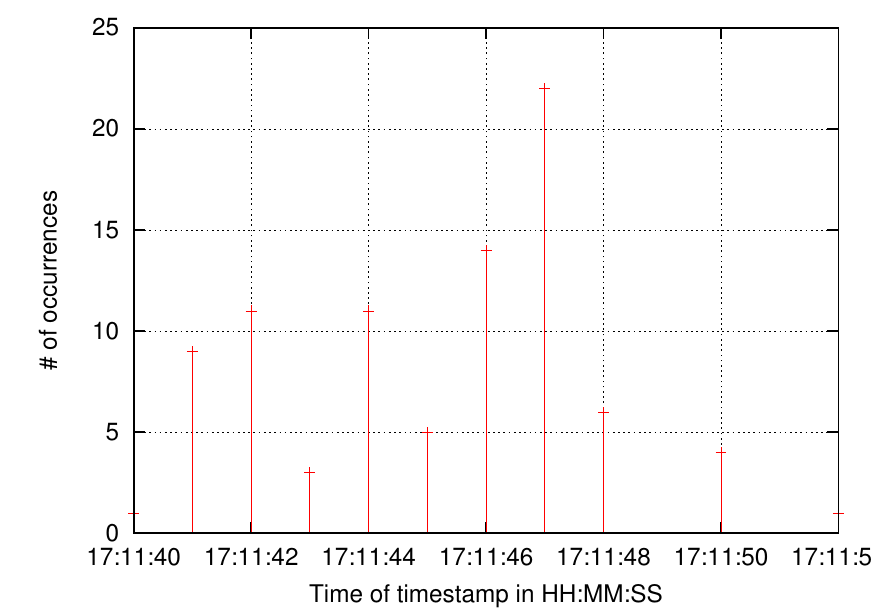}
\caption{Example distribution of computed timestamping times for each cached domain name with a 1 day timestamp.}
\label{fig:timestamp1day}
\end{center}
\end{figure}

\myparagraph{Homogeneity}
We now verify for a given timestamp $T$, if the computed time $t_R$ is the same for all cache entries. Figure~\ref{fig:timestamp1day} shows an example distribution of computed timestamping times for each individual cache entry with a timestamp composed of 100 cache entries. We notice that the large majority of times hold in the interval between 17h11min41sec and 17h11min50sec. Thus the accuracy of the timestamp is around $9$ seconds. 

While Figure~\ref{fig:timestamp1day} depicts a typical distribution, outliers in the distribution of computed  timestamping times of individual cache entries can occur. 
In a few cases, $<< 1\%$ of cache entries, \DNStamp\ computed timestamping times far before or after the actual timestamping time. 
We identified two causes for these outliers: 
(i) the domain name was already cached by the resolver when requesting the timestamp and its TTL is not reset by the timestamp request; the verification process computes a timestamping time prior to the actual timestamping time for this particular cache entry.
(ii) the authoritative server returns a wrong reference TTL during the verification process. 
We observed that a small number of authoritative servers do not implement DNS caching correctly. An authoritative server should always return the reference TTL when receiving a domain resolution request (\domain{A} query) for the domain it is authoritative for. Yet, we noted that a small fraction of authoritative servers does not conform to this behavior and instead decrements the returned TTL as a normal resolver. Thus, the verification process computes a timestamping time prior to the actual timestamping time for this particular domain name.

\myparagraph{Measurements from several locations}
We performed additional experiments to verify whether timestamps can be requested and verified from different locations in the Internet. The different locations that we use for our tests are: (i) Amazon cloud, by running the requesters and verifiers on EC2 instances (ii) broadband Internet access, by testing two standard ADSL lines provided by two different ISP providers.
The tests consist in requesting a different timestamp on each of these locations and verify the generated timestamps from all other locations. We observe no significant differences between the different locations. All locations can verify the timestamps requested on the other locations. The number of retrieved cache entries is similar from one location to another and in line with the observations of Figure~\ref{fig:timestamp1day}.
We observe that the calculated timestamping time for a specific cache entry has at most 1 second difference from one location to another. This difference may be explained by the network delays for some DNS replies or responses.

\subsection{Overwriting an existing timestamp}
\label{sec:overwrite}

\begin{figure*}
\begin{center}
\subfigure[]{
\includegraphics[width=\columnwidth]{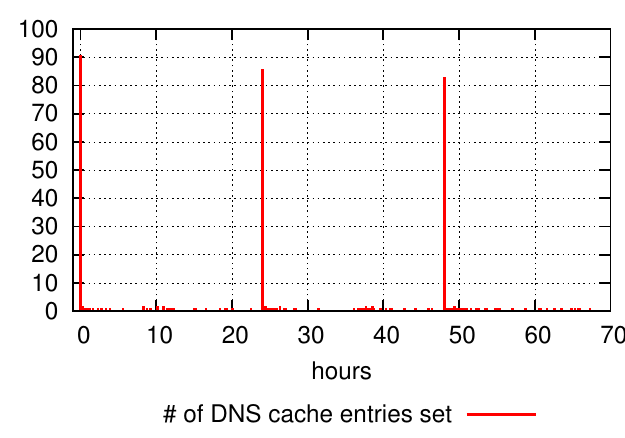}
\label{fig:overwriteTS}
}
\subfigure[]{
\includegraphics[width=\columnwidth]{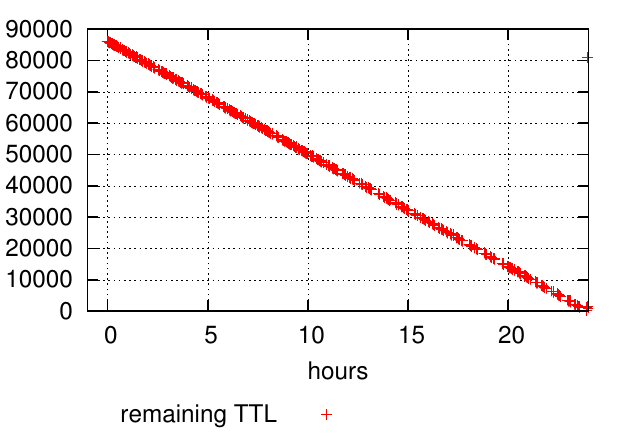}
\label{fig:overwriteTTL}
}
\caption{\subref{fig:domainGen2day}  Cache entries successfully set to the reference TTL during 70 hours continuous overwriting of an existing timestamp. \subref{fig:overwriteTTL} Remaining TTL representing $98\%$ of measured cache entries during the validity period (24 hours) of the initial timestamp. }
\end{center}
\end{figure*}

In this subsection, we experimentally demonstrate that a weak adversary is not capable of overwriting the remaining TTL of an existing timestamp.
We conduct the following experiment to demonstrate this property. We first generate a valid 1 day timestamp consisting of 100 cache entries. Then, we continuously try to overwrite the existing timestamp during 70 hours. We verify whether we can set cache entries to the reference TTL and monitor the remaining TTL each time we try to overwrite a cache entry.

Figure~\ref{fig:overwriteTS} shows the number of cache entries that could be successfully set to the reference TTL. At $t=0$ and at multiples of 24 hours between 80 and 90 cache entries could be updated. This is normal, since this corresponds to the initial timestamp request (at $t=0$) and the subsequent expirations of the timestamp every 24 hours. At these times, the DNS caches do not cache the requested domain names or has just reached a remaining TTL of 0.

Between these peaks and from time to time, the weak adversary is able to set one or two cache entries. This is possible because the domain name selection process encountered some error such as a timeout while reversing an IP address or while requesting the reference TTL. The adversary will thus select a list of domain names different from the list selected by the requester.
This has only a limited impact on the verification. A verifier relies on the domain names and resolver included in $T$ and will not use the cache entries set by the weak adversary. Even if the verifier regenerates the domain names, it is very unlikely that she will select a majority of cache entries the weak adversary has set.
We verified the latter assertion by continuously executing the domain name generation process during the validity period (24 hours) of the initial timestamp and checking the remaining TTL.
Figure~\ref{fig:overwriteTTL} shows that $98\%$ of the remaining TTL decrement exactly as expected. Thus, the attacker cannot modify these cache entries.

\subsection{Timestamps having duration different than one day}
\label{sec:moredays}

We describe experimental results using timestamp durations of 2 days and one hour.
The domain selection process works exactly as described in Section~\ref{sec:domainselection}. The only difference concerns the reference TTL. We accept reference TTL with a value greater or equal to the desired duration.

Our experiments show that it is possible to use \DNStamp\ for timestamps having duration different than one day. Yet, the verifier needs to eliminate falsely computed timestamping times due to DNS cache resolver that set a wrong reference TTL.

\begin{figure*}
\begin{center}
\subfigure[]{
\includegraphics[width=\columnwidth]{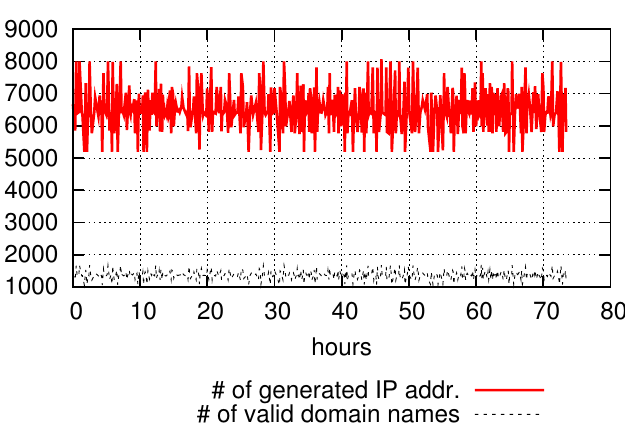}
\label{fig:domainGen2day}
}
\subfigure[]{
\includegraphics[width=\columnwidth]{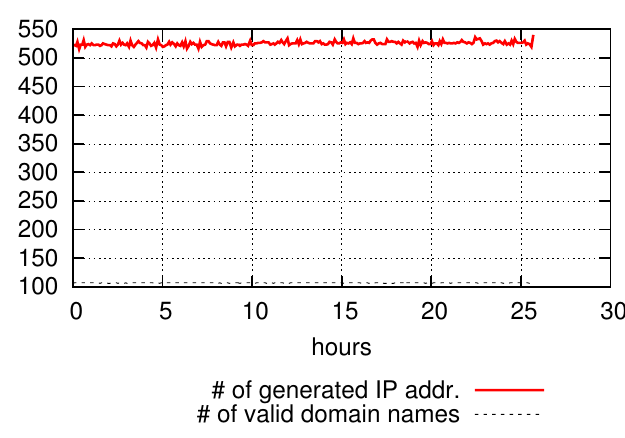}
\label{fig:domainGen1hour}
}
\subfigure[]{
\includegraphics[width=\columnwidth]{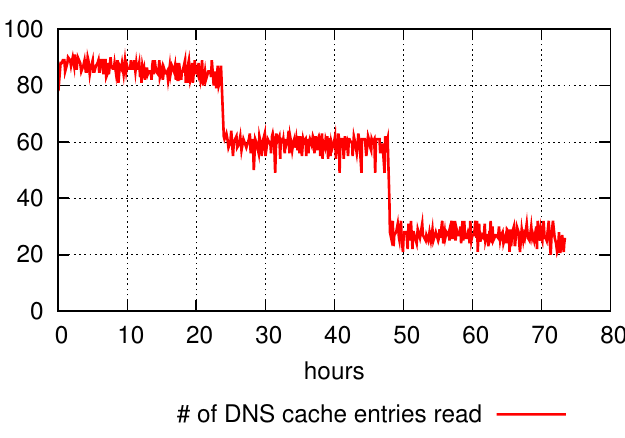}
\label{fig:ReadTS2day}
}
\subfigure[]{
\includegraphics[width=\columnwidth]{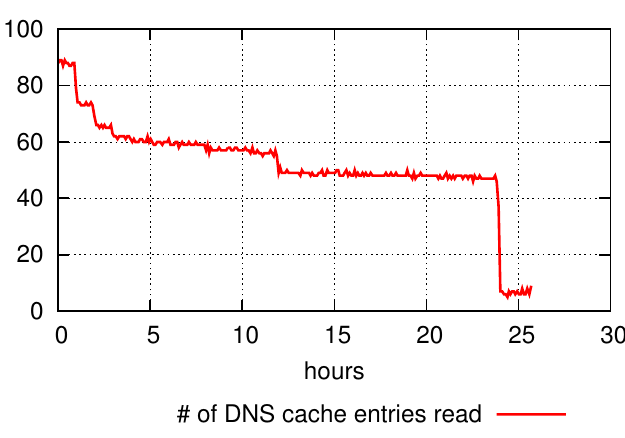}
\label{fig:ReadTS1hour}
}
\caption{\subref{fig:domainGen2day}, \subref{fig:domainGen1hour}:  Number of IPv4 addresses and valid domain names retrieved per verification in order to generate 100 valid domain names with a reference TTL greater or equal to \subref{fig:domainGen2day} two days and \subref{fig:domainGen1hour} one hour. \subref{fig:ReadTS2day}, \subref{fig:ReadTS1hour}:  Number of DNS cache entries read with a timestamp of \subref{fig:ReadTS2day} two days and \subref{fig:ReadTS1hour} one hour. Numbers are averages from 5 two day timestamps $T$ continuously verified during 72 hours, and from 5 one hour timestamps $T$ continuously verified during 26 hours.}
\end{center}
\end{figure*}

\myparagraph{Requesting and verifying timestamps}
We generate 5 two day timestamps which we continuously verify during 72 hours. Similarly, we generate 5 one hour timestamps which we continuously verify during 26 hours.
Figure~\ref{fig:domainGen2day} and  Figure~\ref{fig:domainGen1hour} shows the number of IP addresses and the number of domain names that the domain name generation process has to generate for each verification. 
With a two day timestamp (Figure~\ref{fig:domainGen2day}), about 6500 random IP addresses and 1200 domain names have to be generated, in order to end up with 100 domain names with a reference TTL greater or equal to two days. This confirms the distribution of reference TTL shown in Figure~\ref{fig:distribTTL} where approx.\ $91\%$ of domains have a reference TTL smaller than two days. With a one hour timestamp (Figure~\ref{fig:domainGen1hour}), about 525 random IP addresses and 110 domain names have to be generated.

The high number of IP addresses and domain names required for a two day timestamps can seem prohibitive. Indeed, we require about 30 minutes to request a two day timestamp. In contrast, we only require about 1 minute 30 seconds for the one hour timestamp. We believe that this delay mainly comes from our unoptimized implementation. The optimizations already discussed for one day timestamps should considerably decrease the timestamping delay.

Verifying a timestamp does not necessarily require to reiterate through all IP addresses and domain names.
Instead of executing the whole domain name generation process, as in our experimentation, the verifier may directly use the domain names and DNS cache resolvers indices of $T$. Thus, verification is almost immediate.

\myparagraph{Persistence of timestamps}
Figure~\ref{fig:ReadTS2day} and Figure~\ref{fig:ReadTS1hour} show that the timestamp continues to exist even when the desired reference TTL has exceeded. In the case of two day timestamps, approx.\ one third of the DNS cache entries continue to exist after 48 hours. This portion corresponds to the domain names with a reference TTL strictly greater than two days. Similarly, with one hour timestamp we observe that approx.\ $50\%$ of the DNS cache entries last for one day, reflecting the reference TTL distribution of Figure~\ref{fig:distribTTL}.

To avoid this behavior, it is possible to force reference TTL to exactly the desired duration, resulting in all DNS cache entries to disappear after this duration.

\myparagraph{Errors due to DNS cache resolvers ignoring reference TTL greater than one day}
We observed that some DNS cache resolvers set the remaining TTL to maximum one day, even if the reference TTL is greater than one day. Our experiments show that this represents about one third of the DNS cache resolvers. For instance, Figure~\ref{fig:ReadTS2day} shows that one third of the returned timestamps disappear after 24h, even if the desired reference TTL is greater or equal to two days. This indicates that the remaining TTL for these DNS cache resolvers was set to one day only, and that the remaining TTL reaches $0$ after 24 hours.

\begin{figure*}
\begin{center}
\subfigure[]{
\includegraphics[width=\columnwidth]{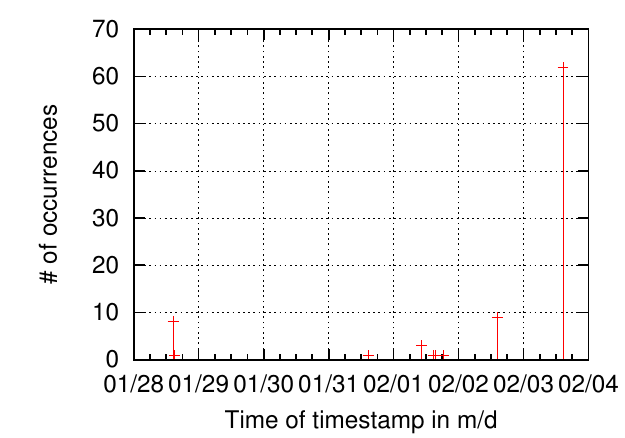}
\label{fig:timestamp2days}
}
\subfigure[]{
\includegraphics[width=\columnwidth]{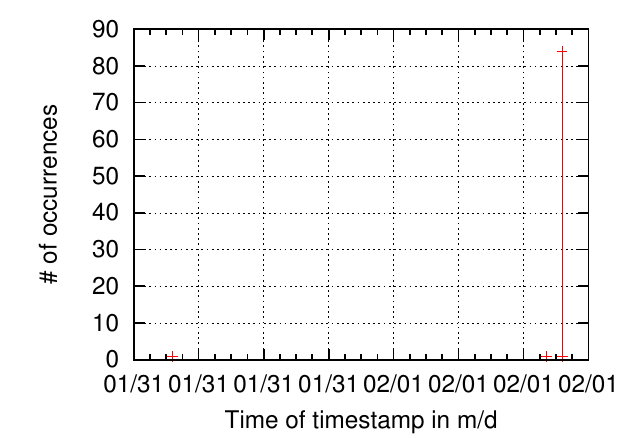}
\label{fig:timestamp1hour}
}
\caption{Example distribution of computed timestamping times for each cached domain name with \subref{fig:timestamp2days} a 2 day timestamp and \subref{fig:timestamp1hour} a one hour timestamp.}
\end{center}
\end{figure*}

\myparagraph{Homogeneity of the computed timestamping times for the different DNS cache entries}
Figure~\ref{fig:timestamp2days} shows the distribution of computed timestamping times for the different DNS cache entries with one example of 2 day timestamp. We notice that the majority of the computed times tends towards the original timestamping time which is Feb 4th at 14h33. Some peaks appear before that date, e.g. at one day and at 6 days prior to the actual timestamping date. Similar behavior holds for one hour timestamp as depicted in Figure~\ref{fig:timestamp1hour}. The outliers represent about one third of the DNS cache entries in the two day timestamp, and only a very small fraction for one hour timestamp.

The outliers can be explained by the fact that some DNS cache resolvers set a false remaining TTL when a new DNS cache entry is added. If the reference TTL of the cached domain name is 7 days and the DNS cache resolver only sets a remaining TTL of one day, the computed time is 6 days prior to the actual timestamping time.
In order to make the timestamping scheme robust to these errors, a verifier must eliminate these outliers by selecting only the most frequent timestamping times.

\section{Related Work}
\label{sec:related}
Trusted timestamping generally relies on a \emph{timestamping service}, provided by a third party centralized service.
The third party receives a document or a document digest, binds the received data with the present time, e.g. by hashing received data together with the present time or by simply appending the present time and signing the results with the private key of the timestamping service. This timestamp can be verified using the public key of the timestamping service.
These mechanisms are standardized \cite{Adams2001,X995}. They rely on entire trust of the third party. Yet, a dishonest third party may generate false timestamps.

Haber and Stornetta \cite{Haber1991} propose two trusted timestamping schemes. In both schemes, the trust is distributed among the participants of the timestamping scheme.
In the first scheme, the timestamping service builds a sequence of timestamps by linking each timestamp with the preceding timestamp. The timestamping service applies a one-way hash function on the digests and on the times of the preceding and of the current timestamps. The timestamping service signs the resulting timestamp and distributes it to the requesting participants. Verification of a timestamp consists in: i) checking the time in the timestamp, ii) asking other participants for the preceding and following timestamps in the sequence and iii) verifying the correct linking and times of these related timestamps.
This reduces the risk of false timestamps produced or inserted in an existing sequence of timestamps.
Haber and Stornetta \cite{Haber1991} state that circumventing this mechanism requires the timestamping service to collude with a significant number of participants. 
Despite this objective, Just \etal \cite{Just1998} show that a collusion attack between a single participant and the timestamping service exists allowing for backdating of timestamps. Just \etal propose several methods for preventing this attack.


The second scheme of Haber and Stornetta \cite{Haber1991} does not rely on any timestamping service. Timestamping is completely distributed among participants.
A participant requesting to timestamp a document, uses the document digest as a seed for a pseudo-random generator. The pseudo-random generator returns a set of participant IDs. The requesting participant sends the timestamping request to each of the returned participants.  Each participant responds with a signed timestamp.
Verification consists in i) checking whether all returned timestamps are properly signed and include the same time and ii) verifying that the pseudo-random generator seeded with the document digest returns the same participant IDs.
This scheme requires all participants to be online and their clocks synchronized. Only a large set of colluding participants would enable breaking the scheme.

Other schemes not relying on a timestamping service rely on broadcast-based timestamping \cite{Benaloh1992}. Broadcast-based timestamping relies on k-ary hash trees that represent all the documents to be timestamped during one timestamping round. Each tree leaf represents a document belonging to a participant. Each leaf is linked to the leaf of the preceding timestamping round. The scheme assumes that the digest of all leaves is broadcasted to all participants. This construction allows to verify whether a given released document did or did not exist in a given timestamping round. Just \etal \cite{Just1998} describes some shortcomings of \cite{Benaloh1992} and propose methods to rectify these shortcomings.

The P2P digital currency system BitCoin \cite{Nakamoto} timestamps the transactions.
BitCoin operates without any central authority and thus requires a completely distributed timestamping scheme. BitCoin relies on linked timestamping and widely publishes the generated hashes in newsgroup or Usenet posts. Each new hash relies on the previous one that has already been published.
A timestamp also includes a proof-of-work, a moderately hard puzzle that the peers of BitCoin try to solve. Once a peer succeeded in solving the puzzle, the peer is rewarded with newly created coins.
Based on BitCoin, CommitCoin \cite{Clark2011} proposes a commitment protocol that enables the sender to prove to a receiver that his commitment existed prior to a time $t$. CommitCoin generates a small BitCoin transaction once the sender committed  its message. As a consequence, the receiver is able to verify and carbon date the commitment.

Schemes not providing timestamping but related to this work were proposed by Ephemeral Publishing \cite{Castelluccia2011}. Their purpose is to allow withdrawing of user-owned content from the internet. User can store ephemeral cryptographic keys, by forcing the insertion of domains in resolvers. The entries are automatically removed from the cache once the TTL has expired. Therefore, the cryptographic key automatically disappears after a delay defined by the user.

\section{Conclusions}
\label{sec:conclusion}

In this work we proposed a new trusted timestamping scheme, called \DNStamp, that exclusively relies on the Domain Name System.
\DNStamp\ does not require a dedicated trusted service nor any form of collaboration among participants using the timestamping service. 
\DNStamp\ can be used without registration to any dedicated service. Thus, anyone with Internet access can request and verify timestamps.


We formalized the security requirements for timestamping and the associated adversarial model. We analyzed the security of \DNStamp\ with respect to this model.
In particular, we showed the resistance to forward-dating and back-dating attacks.
We implemented a command-line tool capable of setting and verifying timestamps in the Domain Name System.
Our experiments showed that we can set and reliably verify timestamps during the validity period of the timestamp.
The experiments also showed that the adversaries with reasonable capabilities cannot overwrite an existing timestamp.

Further work includes extending the validity period of \DNStamp.
This may be achieved by using sequences of linked timestamps or by asking timestamp renewals to a server. These methods may however break our requirement of having no single point of trust.
Finally, we work on an optimized implementation that increases the precision of a timestamp and reduces the requesting time thanks to a higher degree of multi-threading and the use of several inverse resolvers.

\section*{Acknowledgment}
We thank Augustin Soule and Gilles Guette for their insightful comments that helped us improve this paper. We also thank Augustin Soule for providing us the lab DNS workloads.

\bibliographystyle{abbrv}
\bibliography{biblio}

\end{document}